# SiO masers in TX Cam

## Simultaneous VLBA observations of two 43 GHz masers at four epochs

Jiyune Yi[1], R.S. Booth[1], J.E. Conway[1], and P.J. Diamond[2]

[1] Onsala Space Observatory, Chalmers University of Technology, S-43992 Onsala, Sweden

[2] University of Manchester, Jodrell Bank Observatory, Macclesfield, Cheshire SK11 9DL, United Kingdom

**Abstract.** We present the results of simultaneous high resolution observations of $v = 1$ and $v = 2$, $J = 1 - 0$ SiO masers toward TX Cam at four epochs covering a stellar cycle. We used a new observing technique to determine the relative positions of the two maser maps. Near maser maximum (Epochs III and IV), the individual components of both masers are distributed in ring-like structures but the ring is severely disrupted near stellar maser minimum (Epochs I and II). In Epochs III and IV there is a large overlap between the radii at which the two maser transitions occur. However in both epochs the average radius of the $v = 2$ maser ring is smaller than for the $v = 1$ maser ring, the difference being larger for Epoch IV. The observed relative ring radii in the two transitions, and the trends on the ring thickness, are close to those predicted by the model of Humphreys et al. (2002). In many individual features there is an almost exact overlap in space and velocity of emission from the two transitions, arguing against pure radiative pumping. At both Epochs III and IV in many spectral features only 50% of the flux density is recovered in our images, implying significant smooth maser structure. For both transitions we find that red- and blue-shifted masers occur in all parts of the rings, with relatively few masers at the systemic velocity. Thus there is no evidence for rotation, although the blue-shifted masers are somewhat more prominent to the west. At all four epochs red-shifted components are generally brighter than blue-shifted ones. Blue-shifted masers become very weak at some stellar phases but never completely disappear. At Epochs III and IV, we see many filamentary or spoke-like features in both $v = 1$ and $v = 2$ masers, especially in the red-shifted gas. These spokes show systematic velocity gradients consistent with a decelerating outward flow with increasing radius. We outline a possible model to explain why, given the presence of these spokes, there is a deficit of maser features at the systemic velocity. The breaking of spherical symmetry by spoke-like features may explain the high-velocity wings seen in SiO maser single dish spectra.

## 1. Introduction

SiO masers are found in circumstellar envelopes (CSEs) of oxygen-rich, late-type stars on the Asymptotic Giant Branch (AGB) of the HR diagram. The study of CSEs is important because of the high mass-loss rates of their central stars with the ejection of dust and gas, which forms an important source of chemical enrichment of the interstellar medium. They are also of interest because of the diverse astrophysical processes taking place, such as maser excitation, formation of solid state particles, molecular chemistry, and radiation-driven hydrodynamics (Habing 1996). An understanding of how all these processes work in an evolving stellar atmosphere is necessary for a full physical description of the late stages of evolution of low- and intermediate-mass stars.

Circumstellar SiO masers can serve as a powerful tool to study the extended stellar atmosphere/innermost region of CSEs where propagating shock waves, caused by stellar pulsations, play a critical role. The masers arise from rotational transitions of excited vibrational states of SiO and the characteristic temperature of the first excited vibrational level, 1800 K, requires the SiO maser zone to lie very close to the parent star. The outer boundary of SiO masers is limited by the availability of SiO in the gas phase before its condensation on to dust particles. SiO masers observed by Very Long Baseline Interferometry (VLBI) are tracers of the physical conditions and dynamics on scales of only a few stellar radii; scales which cannot be directly imaged by any other astronomical technique. Early VLBI observations using three stations in Europe showed the $v = 1$ 43-GHz SiO masers to have compact structures (Colomer et al. 1992) and the wider range of baselines in the Very Long Baseline Array (VLBA) enabled Diamond et al. (1994) to make the first SiO maser images showing rings of components around the Mira variables TX Cam and U Her. Subsequent observations have confirmed the location of SiO masers within a few stellar radii of the stars in ring-like structures (Greenhill et al. 1995; Boboltz et al. 1997; Desmurs et al. 2000; Yi et al. 2002; Diamond and Kemball 2003; Cotton et al. 2004).

The precise nature of the pumping mechanism for circumstellar SiO masers remains open and this uncertainty is an obstacle to using the SiO masers to interpret physical conditions in CSEs. The radiative pumping models (e.g. Kwan & Scoville 1974; Bujarrabal 1994a, 1994b), in their standard form, predict that strong $J = 1 - 0$ masers in the $v = 1$ and $v = 2$ states cannot occur in a same spatial zone. This constraint is emphasized in the work of Lockett & Elitzur (1992). VLBI observations with a resolution of 7 mas by Miyoshi et al. (1994) showed maser spots in the two lines to be coincident within 2 mas and therefore argued for collisional pumping. In retrospect, it is apparent that their resolution was insufficient to decide about the spatial coincidence of the $v = 1$ and $v = 2$ masers since this is much larger than the size of the maser spots themselves (down to sub-milliarcsec) and is a significant fraction of the ring radius. Subsequent (single epoch) higher resolution observations using the VLBA ($0.7 \times 0.2$ mas) by Desmurs et al. (2000) showed the $v = 2$ masers were found systematically to be on rings of smaller radius. This result was used to argue for the radiative pumping hypothesis. We note however that this result is based on measurements at a single epoch (see Sect. 5.1).

Model calculations by Humphreys et al. (1996) with SiO masers pumped predominantly by collisions also show $v = 2$ masers lying a little inside the equivalent $v = 1$ masers. Further work, by Gray & Humphreys (2000), based on the same pumping model, predicted that the ring radius of the two masers and the displacement of the two maser rings varies over stellar phase. It has become clear that the spatial coincidence, or not, of the maser in the two lines observed at one epoch cannot be used as a conclusive argument in favour of one pumping mechanism over another. Instead, multi-epoch observations of the relative positions of the two transitions are required. Such observations, together with detailed modelling, can hopefully constrain the dominant pumping scheme and therefore shed light on physical conditions in extended stellar atmospheres.

This paper reports the results of multi-epoch monitoring of the $v = 1$ and $v = 2$ lines in TX Cam. Our study differs from other previous work comparing these transitions in two respects. First we observed at several epochs, sampling a large fraction of the stellar pulsation cycle. Secondly we used a special technique to try to determine relative positions of the two transitions. Previous observations, although simultaneously observing both transitions, effectively reduced the data from each transition separately using self-calibration techniques, losing relative position information. To try to recover the correct registration these previous studies then identified SiO features in the two transition maps which had similar velocities and spatial structure. Therefore, the comparison of the locations of the masers in the two lines was subject to the assumption of coincident reference features in the two images. Our alternative technique, (see Sect. 2) monitors the phase difference between the data for each transition and hence seeks to retain information about the relative position of masers in the two transitions.

**Table 1.** Summary of the characteristics of TX Cam

| Parameters | Value[*] |
|---|---|
| Other identifiers | IRAS 04566 + 5606, IRC + 60150 |
| Spectral type | M8 – M10 |
| Variable type | Mira |
| Pulsation period | 557.4 days |
| Distance[a] | 380 pc |
| Radial velocity[b] | 11 km s$^{-1}$ |
| Masers from other molecules[c] | H$_2$O & OH, absent |

[*] Unmarked values are from the SIMBAD data base
[a] Olofsson et al. (1998)
[b] From the SIMBAD data base but since the velocity differs from molecule to molecule and due to the variability of SiO masers, different values can be found in other references. Some references are, 9.6-12.5 km s$^{-1}$ from CO by Olofsson et al. (1991); 11.8 km s$^{-1}$ from CO, 9.8 km s$^{-1}$ from SiO ($v = 1, J = 2 - 1$) by Cernicharo et al. (1997); mean velocities of 10.7-11.0 km s$^{-1}$ from SiO ($v = 1, 2, 3, J = 1 - 0$) by Cho et al. (1996)
[c] Benson et al. (1990); Benson & Little-Marenin (1996); Lewis (1997)

Our first observations were made of the Mira variables TX Cam and R Cas using the VLBA operated by the National Radio Astronomy Observatory (NRAO)[1] in 1998. In this experiment, the maser ring in TX Cam was significantly disrupted in both maser lines, while R Cas showed a clear ring-like structure of the masers. These differences were almost certainly due to the different stellar phase at which each source was observed (Yi et al. 2002). In both sources the $v = 2$ maser distribution occurred at slightly smaller radius than the $v = 1$ masers. After the success of this initial experiment multi-epoch VLBA observations of both sources were proposed and executed. In this paper we present the results of the observations toward TX Cam at four epochs. The observational results of R Cas will be reported in a future paper. General characteristics of TX Cam are given in Table 1.

The paper is organized as follows. In Sect.2 we give a general description of our observations. In Sect.3 we present our data calibration scheme including a description of the phase referencing scheme devised to measure the relative positions of the two maser transitions. In Sect.4 observational results and the analysis of the maser ring properties are given. In Sect.5 we discuss our results, comparing with other observations and the predictions of maser models. The summary and conclusions are given in Sect.6.

## 2. Observations

The results of VLBA observations of TX Cam at four epochs spanning a stellar cycle are described in this paper. Table 2 lists the dates of observation, experiment duration and the corresponding stellar phase as defined by the visual light curve (see Fig. 1). The first observation was made in 1998 and the latter three in 2000/2001. These last three epochs were scheduled approximately four months apart in order to try to investigate stellar phase-dependent properties of the $v = 1$ and $v = 2$ lines. For TX Cam a four month interval corresponds to a difference in the stellar phase of ∼0.2 cycle. In all the epochs except Epoch II all 10 VLBA antennas were used. During Epoch II the Los Alamos antenna was not available due to a nearby forest fire; this loss significantly reduced the short baseline coverage of the array at this epoch. The Epoch II observations also suffered from a number of intermittent losses from other antennas.

At each epoch we used a special observing setup and schedule to try to obtain accurately the relative positions of masers in the two SiO maser lines. The frequency band was divided into eight separate IF channels, (all right circular polarisation), each with a bandwidth of 8 MHz (covering ∼ 56 km s$^{-1}$ in velocity). The lowest frequency IF channel (IF1) was centred on the Doppler-shifted frequency of the $v = 2$, $J = 1 - 0$ line (rest frequency 42.820542 GHz) while the highest frequency channel (IF8) was placed at a centre frequency to cover the $v = 1$, $J = 1 - 0$ line (rest frequency 43.122027 GHz). These two IFs will be referred to subsequently in the paper as the $v = 2$ and $v = 1$ IF channels. The other six IF channels were distributed evenly in frequency over the 301 MHz between the two maser transitions. The observing sequence switched rapidly between TX Cam and a continuum calibrator and continuum signals were detected in all eight IF channels. By fitting the group delay across all eight IF channels (phase derivative versus frequency) during these continuum observations the relative phase between IF1 and IF8 could be determined. This relative phase was then interpolated in time to the TX Cam observations allowing the relative phase of the spectral line visibilities in the IF1 and IF8 channels to be calibrated. This in turn then allowed the relative positions of $v = 1$ and $v = 2$ features to be determined.

A relatively bright continuum calibrator was required to allow fringe detection. The nearest suitable calibrator we could find in 1998 was J0359+5057 which was 9° away from TX Cam. This calibrator was observed in a repeated sequence of a 4 minute scans followed by 7 minute scans on TX Cam. The data were correlated at the VLBA correlator in Socorro, New Mexico where auto- and cross-correlation spectra were produced with 256 spectral channels per IF band corresponding to a ∼0.2 km s$^{-1}$ channel spacing.

As described in Sect. 3.3 below, our technique for cross-calibrating the $v = 1$ and $v = 2$ lines was only partially successful. Despite this we give in Sect. 3 a full description of the technique and data analysis because we believe it will be very useful to others attempting this method. With some modifications we believe accurate relative positions of the masers in the two transitions can be achieved using the technique (see Sect. 3.4).

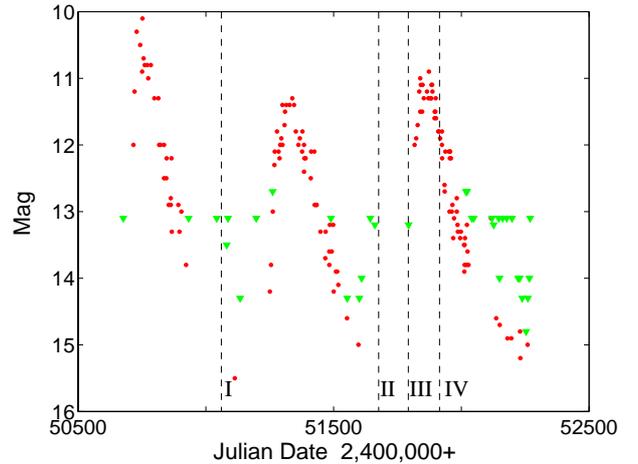

**Fig. 1.** Stellar optical light curve of TX Cam (courtesy of AAVSO). The vertical dashed lines in the plot represent the times of the four epochs of VLBA observations. The downward triangles indicate upper limits to the optical magnitudes.

## 3. Data reduction

### 3.1. Calibration

All data reduction was done using the NRAO AIPS package. We first applied standard calibration steps for bandpass, Doppler shift, and amplitude. For phase calibration special techniques were needed to allow the estimation of the relative $v = 1$ and $v = 2$ maser positions.

Bandpass calibration was determined using the continuum calibrator data. AIPS task CPASS first used the autocorrelations to solve amplitude response of the antenna bandpasses. The cross-power spectra on the calibrator were then used to estimate the phase responses of the bandpass filters. Using the task CVEL, Doppler corrections were made to the data to remove the effects of motion of the VLBA antennae relative to the Local Standard of Rest.

Amplitude calibration of the data was achieved using the total-power spectra of TX Cam based on the "template spectrum" method. Here a template total power spectrum in Janskies (Jy) was determined. Then the conversion factor between correlation coefficients and Jy for each antenna as a function of time was found by comparing this template spectrum with the time-variable antenna autocorrelation spectra (measured in correlation coefficients). Finally these time-variable station-based conversion factors are used to correct the cross power visibility data. The template spectra at each epoch were obtained using high elevation observations. The conversion factor was obtained from the recorded system temperature in Kelvin multiplied by the antenna gain in Jy/K. Because of the high elevations used for computing the templates and the relatively low system temperatures (∼ 90K) we expect that opacity effects on our templates are relatively small and that our absolute amplitude scale is accurate to better than 20%. Relative gains achieved by this method are typically < 5%.

The first step in phase calibration was to determine the time constant phase offsets introduced at each IF by the electronics. This was done using a single 2 minute scan on calibra-tor J0359+5057 and separately fringe fitting each of the 8 IF bands. The resulting fringe rates were set to zero and the constant phase offsets between IFs removed from all the data. After this step there remained the effects of the atmosphere and unmodelled array geometry which causes a change in residual delay versus time ($\tau(t)$). Residual delay equals the derivative of phase with respect to frequency and causes a time variable phase ($\phi = \nu\tau(t)$) at each IF and a corresponding phase difference between IFs ($\Delta\phi = \Delta\nu\tau(t)$). The time variable delay was estimated by fringe fitting all IFs of the continuum data solving simultaneously for phase, phase-rate, single- and multi-band delay. The results were then interpolated to the times of the spectral line observations allowing the long term relative phase difference between IF1 and IF8 to be estimated and removed from the data.

**Table 2.** Observations of four epochs

| Epoch | Date | Duration (UT) | Stellar phase[a] | Restoring beam (mas)[b] |
|---|---|---|---|---|
| I | 1998/09/04 | $08:33 \sim 12:30$ | $\phi = 0.52$ | $\sim 0.6 \times 0.4$ |
| II | 2000/05/12, 13[c] | $19:09 \sim 01:00$ | $\phi = 0.60$ | $\sim 1.05 \times 0.6$ |
| III | 2000/09/05 | $10:01 \sim 15:00$ | $\phi = 0.83$ | $\sim 0.4 \times 0.2$ |
| IV | 2001/01/06 | $02:31 \sim 07:30$ | $\phi = 0.05$ | $\sim 0.5 \times 0.2$ |

[a] Stellar phase, $\phi$ varies from 0 to 1 during one cycle. $\phi = 0$ for visual maximum. The stellar phases given here are determined using the same method as that of Diamond & Kemball (2003) with the optical maximum at JD = 2450773.0 and a mean period of 557.4 days.

[b] 1 mas corresponds to $0.5685 \times 10^{11}$ m at the assumed distance to TX Cam.

[c] The Los Alamos antenna was not used.

At 43 GHz significant random delay variations due to the atmosphere occur over fractions of a minute; thus the 10 minute target-calibrator cycle time was not fast enough to track these short term fluctuations. The data therefore remained phase incoherent and application of spectral line self-calibration techniques was needed before imaging the data. We applied such self-calibration to the $\nu = 1$ data set, first identifying a compact spectral feature, then iteratively self-calibrating and imaging this feature. The final antenna-based phase solutions were then applied to all spectral channels in both the $\nu = 1$ and $\nu = 2$ data sets. The transfer of the phase solutions from one IF to the other rather than independent self-calibration of each IF was essential in order to preserve the relative position information between the two transitions.

### 3.2. Imaging and correcting for effects of correlation phase centre error

After calibrating the data we next created image cubes of $\nu = 1$ and $\nu = 2$ masers of four epochs using the AIPS task IMAGR. The size of the synthesised beam of the maps at four epochs is given in Table 2. The relatively large beam size for Epoch II is due to the loss of visibility throughout the calibration processes on the three longer baselines. These losses occurred because at that epoch the continuum calibrator was relatively weak and resolved.

On comparing the relative positions of the $\nu = 1$ and $\nu = 2$ maser features having the same velocity and morphology, we found at each epoch a systematic shift of the order of 2 mas.

It was eventually realised that this was caused by the large difference between the true absolute position of TX Cam and the position used for correlation. Such an absolute position error, of $\Delta\alpha$ in RA, using our data reduction technique will produce a relative error in RA between the $\nu = 1$ to $\nu = 2$ positions of $(\Delta\nu/\nu)\Delta\alpha$, where $\Delta\nu$ is the frequency difference between the two transitions. To see how this arises note that the correlation position error causes a time variable delay error $\tau(t)$ on each baseline, giving a phase error $\phi(t) = \nu\tau(t)$ at each frequency. Since this error occurs only in the TX Cam observations, the continuum delay solutions toward the calibrator did not correct these errors. The subsequent spectral line self-calibration in $\nu = 1$ *did* remove the position phase errors from the $\nu = 1$ data. However, when these phase corrections were transferred to the $\nu = 2$ data they could not fully correct the $\nu = 2$ data because the source position phase error is frequency dependent. After cross self-calibration the residual error in the $\nu = 2$ data was $\Delta\phi(t) = \Delta\nu\tau(t) = (\Delta\nu/\nu)\phi(t)$. This equals the phase error expected due to a shift in the relative $\nu = 1$ and $\nu = 2$ positions of $(\Delta\nu/\nu)\Delta\alpha$.

To correct the relative positions of the $\nu = 1$ and $\nu = 2$ lines we attempted to use fringe-rate mapping (Walker 1981) to find the difference between the correlated and true positions of the $\nu = 1$ data, and then apply a shift of $\Delta\nu/\nu$ times this error to the relative $\nu = 1$ and $\nu = 2$ positions. We applied this method to both the Epoch III and IV data and found absolute position offsets consistent with what was expected if most of the relative position error between $\nu = 1$ and $\nu = 2$ lines were due to a correlator position error. However, there were large uncertainties in the derived positions (up to one third of the error). These uncertainties arise because the fringe-rates are affected by atmospheric effects as well as the correlator position error. We concluded that the fringe-rate mapping method was not accurate enough to correct the relative positions.

### 3.3. Final registration of $\nu = 1$ and $\nu = 2$ data cubes

Because of the difficulties on applying absolute calibration (see Sect. 3.2) the final alignments of the $\nu = 1$ and $\nu = 2$ masers at Epochs III and IV were done by finding common maser features at the two transitions. At both epochs we manually shifted the $\nu = 2$ relative to the $\nu = 1$ image until reference features coincided. The features used for registration are marked with an 'R' in Fig. 2(c), 2(d). These features were chosen be-cause of their similar shapes and velocities in both transitions, but we note that their registration results in alignment of at least six more components in each case, so in principle any pair out of these could have been used. The relative shifts between $v = 1$ and $v = 2$ found at the two epochs were 1.49 mas in RA and 1.98 mas in Dec for Epoch III and 1.26 mas in RA and 1.98 mas in Dec for Epoch IV. The difference in estimated shifts of 0.23 mas between Epoch III and IV could be simply a measurement uncertainty in the manual alignment process, however we believe our accuracy is much better than this. Alternatively the absolute calibration process could really be giving different shifts at the two epochs, either due to atmospheric contributions or source proper motions. Although it can provide some contribution we do not expect that proper motion accounts for most of the 0.23 mas difference. There is no measured proper motion of TX Cam reported in the literature, but if we assume an upper limit to the proper motion velocity of 100 km s$^{-1}$ the resulting source proper motion is 57 mas/yr. This then would cause a difference in the $v = 1$ to $v = 2$ shift between Epochs III and IV of only 0.13 mas. We favour instead that most of the 0.23 mas difference in RA shift between the two epochs is introduced by the atmosphere. Supporting this conclusion are the large errors in estimating the absolute position of the targets using fringe-rate mapping (see Sect. 3.2) which show that residual atmospheric phase errors must be large.

For Epochs I and II there are many fewer maser features in total and even fewer features in common between the two transitions and so manual alignment was not possible. For these epochs we used the absolute position calibration but applied the $v = 1$ to $v = 2$ position shift found at Epoch IV to account for the correlator position error. We can provide a crude estimate of the accuracy of the relative $v = 1$ to $v = 2$ position accuracy for Epochs I and II if we suppose it to be comparable to the difference in shift between Epoch III and IV, i.e. of order 0.2 mas.

For future observations of TX Cam it will obviously be useful to use a more accurate correlation position. This is especially important if our technique for aligning $v = 1$ and $v = 2$ is used (see next section). From our data we can produce an estimate of the true position of TX Cam relative to our correlator position if we take the average of the manual shifts required to align $v = 1$ and $v = 2$ maps at Epochs III and IV and multiply these by $v/\Delta v$. The resulting shift is consistent with the lower accuracy fringe-rate mapping of the $v = 1$ data described in Sect. 3.2. The final position estimate for the shell centre at epoch 2000.9 is then $\alpha_{J2000} = 05^h00^m51^s160$ and $\delta_{J2000} = 56°10'54''.046$. The accuracy of this position is hard to estimate. If we consider the 0.2 mas difference in $v = 1$ to $v = 2$ alignment between the two epochs as indicative of the atmospheric error contribution and double this for safety, we estimate that our absolute position has a 50 mas uncertainty. Our new position would then be marginally outside the error bars given by Baudry et al. (1995); the difference might be due to the unknown stellar proper motion, which could be as large as 50 mas/yr.

### 3.4. Future improvements to the technique

Our attempt at calibration of the $v = 1$ and $v = 2$ relative position was only partially successful. It was used primarily for Epochs I and II and even here empirical corrections from Epoch IV were needed to account for correlator position error. However the results in Epochs I and II show the importance of attempting such an absolute calibration since at maser minimum it is very difficult to register the $v = 1$ and $v = 2$ maser maps even approximately using coincidence of features due to the disruption of the maser rings and the lack of clearly coexisting masers in both lines. It would also be convenient to avoid the uncertainties introduced by manual alignment even at epochs with plenty of features. We believe that the technique can be improved to achieve sub-milliarcsecond accuracy in relative position calibration. Our results show the importance of having a good absolute position of the source at the time of correlation. If a good position is not available we should optimise the schedule to do fringe-rate mapping to allow estimation of the absolute position, enabling us to make post-correlation corrections. In addition the different relative shifts found between Epoch III and IV suggests that residual atmospheric errors also limited our accuracy to about 0.2 mas. In order to improve on this a quicker switching cycle and closer calibrator should be used. The 10 minute cycle used could be reduced to say 3 minutes. In addition, now that a denser grid of calibrators at 43GHz has been established, a much closer calibrator for TX Cam could be found.

## 4. Results

### 4.1. General description of the maps by epoch

Maps of the maser emission in the $v = 1$ and $v = 2$ lines at all four epochs are presented in Fig. 2. This diagram shows the total-intensity images, summed over all velocity channels with emission of each transition as superimposed contours with different colours. In the last two epochs the maser structure is much richer and we give detailed information in other figures. In Fig. 3 we present in colour the integrated intensity maps separately for the two transitions at the last two epochs. Figs. 4, 5, 6 and 7 show the velocity field at these epochs in the two lines. Below we give a brief summary of the structures seen in each epoch. In Sects. 4.2 and 4.3 we discuss specific features such as extended flux density or ring diameter as a function of epoch, concentrating on the last two epochs for which we have the best data.

Epoch I (see Fig. 2(a)) has relatively few maser features and it is hard to make out any ring-like distribution. Many maser features are detected in only one of the transitions. To the northeast the $v = 1$ masers seem, in general, to occur at a larger radius than the $v = 2$ masers. Only in the two components to the south-west is it possible to associate features in detail. Here again the $v = 1$ masers reside at a slightly larger radius than their $v = 2$ counterparts (maser clumps with a similar size and shape, occurring within the same velocity range).

Epoch II (see Fig. 2(b)) shows that the $v = 1$ masers form a rough ring shape. The $v = 2$ transition, however, lacks emission to the south-east while in the strong emission complex to the north-east there is considerable overlap of components in both transitions. Although the deficiency of a clear $v = 2$ maser ring makes it difficult to make a strong statement, we estimate that the $v = 2$ masers overall are found in a slightly smaller region, with obvious exceptions.

At Epoch III (see Figs. 2(c) and 3), both $v = 1$ and $v = 2$ masers have developed a clear ring structure, most of the maser features residing within a diameter of $\sim 33$ mas (or 12.5 AU at the distance of 380 pc). Individual maser spots occur in the same zones. Furthermore, in addition to the reference feature, several clumps, especially in the north-east, almost exactly overlap in both spatial structure and velocity (see Figs. 4, 5). Colour total-intensity images, integrated over all velocity channels containing emission, of both $v = 1$ and $v = 2$ masers are shown on the left side of Fig. 3. Becoming prominent in these figures are radially extended features (filaments or spokes) in which the intensity peaks in their central regions; such features are even clearer in Epoch IV. Both maser transitions have such extended complexes co-existing to the north but many small compact spots to the south are seen only as $v = 1$ masers without $v = 2$ counterparts. The maps of the velocity distribution of both masers at this epoch are presented in Figs. 4 and 5. We also show total-power spectra superimposed on the spectra of the summed flux of the maser emission recovered by the synthesized images. The velocity of both masers ranges from $\sim 2.5$ km s$^{-1}$ to $\sim 19$ km s$^{-1}$. Assuming the stellar velocity of 11 km s$^{-1}$, the blue-shifted $v = 1$ masers are as bright as the red-shifted ones. On the other hand the $v = 2$ masers are dominated by the red-shifted clumps and blue-shifted components below 5 km s$^{-1}$ are almost absent. There is no clear signature of an ordered motion such as rotation in the velocity field but the blue-shifted $v = 1$ masers are found predominantly on the west side of the map. In several features and especially in the clump complex to the north-east there is an apparent radial velocity gradient consistent with a decelerating outward flow.

At Epoch IV (see Figs. 2(d) and 3), both maser transitions form a fully developed circular ring and most masers are spatially extended on scales of several milliarcsec. The $v = 1$ masers are more homogeneously developed around the ring while the $v = 2$ masers show a partial maser-free region on the south-east area of the map. Total-intensity images presented in the right column of Fig. 3 show many spoke-like features, which also contain local brightness enhancements. Most masers lie within the diameter of $\sim 36$ mas (or 13.7 AU) in both transitions but maser clumps to the north and north-east extend beyond this radius. Both transitions show a ring thickness of several milliarcsec but the $v = 2$ maser ring is slightly thinner (see Sect. 4.3 for details). There are many maser features that exist in both transitions and completely or partially overlap in both space and velocity. For the spoke-like clumps in the north-west and south-west parts of the ring the $v = 1$ masers extend out to a larger radius than the $v = 2$ masers. Figs. 6 and 7 also show the maps of velocity distributions of the two maser transitions at this epoch. The $v = 1$ maser emission appears from $\sim 1.8$ km s$^{-1}$ while the $v = 2$ maser starts at $\sim 2.8$ km s$^{-1}$. Emission disappears at $\sim 19.5$ km s$^{-1}$ for both transitions. Both blue- and red-shifted components are seen in the $v = 1$ transition, but the $v = 2$ emission is dominated by red-shifted emission and the blue-shifted emission is very weak. The blue-shifted masers in the $v = 1$ line occur all around the ring, but the most prominent ones are on the west side of the map. In both transitions, velocity gradients along the spoke-like features indicate a decelerating outward flow velocity with increasing radius.

## 4.2. Extended emission

What is often not recognised is that images of SiO masers obtained by the VLBA do not recover all of the single dish flux density. This implies that a significant fraction of the maser emission is in large angular-sized structures that are resolved even by the shortest baselines of the VLBA. While this emission component could be truly smooth and continuous, it is more likely, given we are observing maser emission, that it is due to the superposition of many weak maser spots. The evidence for missing flux is most clearly seen in data from Epochs III and IV if we compare the spectrum of emission recovered in the cubes with that in the autocorrelation spectra (see Figs. 4, 5, 6 and 7). At most velocities the fraction of recovered flux density is similar for both the $v = 1$ and $v = 2$ transitions and ranges between 50% and 60%. It is interesting to see that for both transitions and both epochs the lowest fraction of recovered flux density occurs at around 9 - 11 km s$^{-1}$. This could be an indication of more extended maser emission around the stellar systemic velocity. Fitting a Gaussian function to the short baseline (including zero spacing) data allows us to estimate the missing flux to be in structures larger than 3 - 4 mas at 10 km s$^{-1}$ at Epoch IV and larger than 4 - 5 mas at Epoch III. At Epoch III the maser emission near 10 km s$^{-1}$ is almost entirely resolved, indicating a lower limit to the size of the extended features of 4 mas.

## 4.3. Maser ring properties

An important parameter for constraining maser models is the relative radii at which the $v = 1$ and $v = 2$ masers occur. Models by Gray & Humphreys (2000) suggest that the relative positions of the two transitions are dependent on the stellar phase. As noted in Sect. 4.1 for Epoch I there is some indication that the $v = 1$ masers lie at larger radii than the $v = 2$ masers while for Epoch II the relative separations were less clear. It is hard to analyse the first two epochs quantitatively because of the relatively few maser features and the disruption of the ring shape, therefore we concentrate on Epochs III and IV where there were well-developed ring structures. At these epochs we used the AIPS tasks IRING to find the radial distribution of the masers. The logarithm of the integrated intensity images was taken first to reduce the effect of dominating bright features on the radial profile.

In order to plot radial profiles the ring centre had to be found and this was estimated by the following iterative procedure. First, a trial centre position was defined by eye. The AIPS task IRING was then run to estimate a radius which en-closed half of the logarithmic flux density (cf. Cotton et al. 2004). Using this radius, slightly different x positions for the ring centre were then tested. The best x position was obtained by finding the location in which the logarithmic flux density inside a semicircle of radius R and azimuth 0 to 180 degrees equaled the logarithmic flux density beyond R with the same azimuth limits. In a similar way the best y position was determined by looking at the flux density inside and outside a semicircle with azimuth limits -90 to 90 degrees.

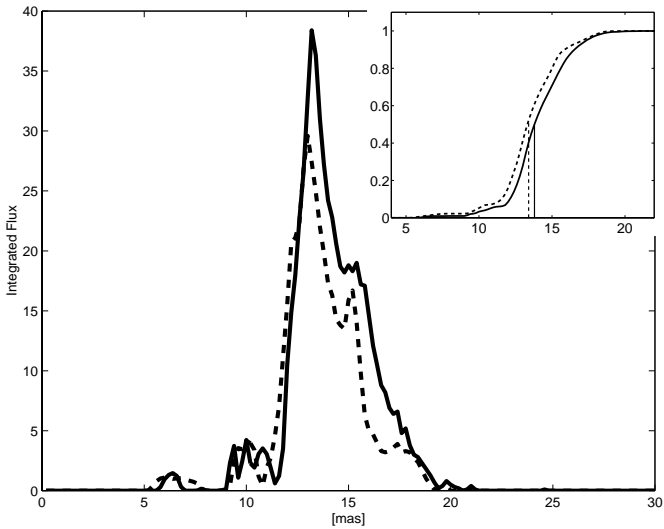

**Fig. 8.** The integrated flux of the v=1 maser emission (solid line) and v=2 maser emission (dashed line) per annulus at Epoch III. The intensity of the maps was logarithmically scaled prior to the analysis. The inset represents the normalised cumulative flux versus radius of the two masers. The vertical lines in the inset indicate the radii of the maser ring containing 50% of the total emission.

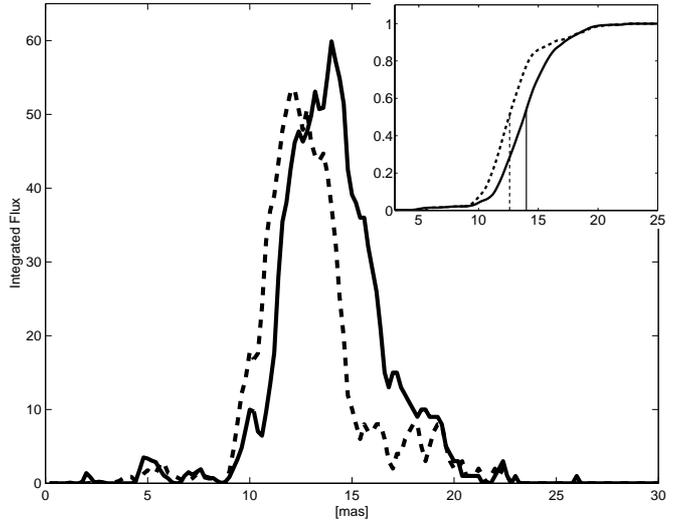

**Fig. 9.** The integrated flux of the v=1 maser emission (solid line) and v=2 maser emission (dashed line) per annulus at Epoch IV. The intensity of the maps was logarithmically scaled prior to the analysis. The inset represents the normalised cumulative flux versus radius of the two masers. The vertical lines in the inset indicate the radii of the maser ring containing 50% of the total emission.

The above method of finding the ring centre was used only on the transition with the clearest ring structure ($v = 1$ maser in both Epochs III and IV). Then radial profiles for the $v = 1$ and $v = 2$ maser using this common centre were plotted. Figs. 8 and 9 show the radial profiles for Epochs III and IV. For Epoch III the bulk of the maser emission at the two transitions is found to be at almost the same radius, the peak of the $v = 2$ maser being displaced by only 0.2 mas inward relative to the $v = 1$ maser peak. It is clear however that the $v = 1$ maser covers a slightly wider range of radii. The mean ring radius and its thickness at each transition was determined using plots of cumulative flux as shown in the inset to Fig. 8; this shows that the radius of the maser ring containing 50% of the total emission is 13.8 mas in the $v = 1$ maser and 13.4 mas in the $v = 2$ maser. The outermost extent of the ring, defined by the radius with 90% of the total emission, is 16.6 mas in the $v = 1$ maser and 16.0 mas in the $v = 2$ maser. If we define the ring thickness as the difference in radii between the cumulative flux ranges at 25% and 75%, then the $v = 1$ maser ring thickness is 2.4 mas compared to the $v = 2$ maser ring thickness of 2.2 mas.

Fig. 9 presents the same analysis made for Epoch IV. Most of the maser emission from the $v = 1$ line is observed over a wider range of radius and is further out than the $v = 2$ emission. The displacement between the two masers is more clearly seen than in Epoch III. The peak positions are offset by 1.8 mas. From the cumulative flux plot we find that the radius containing 50% of the total flux is 14.0 mas for the $v = 1$ maser and 12.6 mas for the $v = 2$ maser. The furthermost extent of the ring, defined as the radius enclosing 90% of the total flux, is 17.2 mas in the $v = 1$ masers and 16.4 mas in the $v = 2$ maser. Defining the ring thickness as between the radii where the cumulative flux changes from 25% to 75%, we find that the $v = 1$ maser ring thickness is 3.0 mas compared to a $v = 2$ maser ring thickness of 2.6 mas.

## 5. Discussion

In this section we discuss our results in the context of other observations of TX Cam, both in general and in terms of the relative positions of the two maser transitions. We also compare our results to the proposed models. In Sect. 5.1 we discuss other observations of TX Cam while in Sect. 5.2 we compare our v=1 and 2 data to theoretical models and maser pumping schemes. In Sect. 5.3 we comment on the regular 'spoke-like' structures seen in intensity and the velocity field and suggest possible explanations.

### 5.1. Comparison with other TX Cam observations

Our observations have tracked the relative positions of the SiO masers in the $v = 1$ and $v = 2$ states of the $J = 1 - 0$ transition over four epochs, spaced across a stellar cycle of TX Cam and have shown changes, not only in the overall distribution of the masers but also in the relative distributions of the emission in the two transitions.

Variations in the $v = 1$ line have been shown previously by Diamond & Kemball (2003; DK03 hereafter) in their $v = 1$,

$J = 1 - 0$ maser monitoring toward TX Cam. A clear agreement with DK03 is seen in the broken ring of their map labelled $\phi = 1.49$ (see Fig. 2 of DK03), which was observed about 1.7 weeks before our Epoch I. This almost exact agreement confirms the observed disruption of the maser ring near maser minimum light. Note that only our Epoch I data lies within the stellar cycle movie presented by DK03; our remaining epochs occur in the subsequent cycle, and so differences in what we observe and what DK03 observes may be due to cycle-to-cycle variations. It is instructive however to compare what we observe in the $v = 1$ transition with what DK03 observe in $v = 1$ at similar phases of the previous stellar cycle. Our results are in general agreement in the sense that clear ring structures are observed near maser maximum, breaking up into less distinct (but still ring-like) structures near maser minimum.

DK03 describe the dominant emission as being confined to a narrow projected ring at all epochs. This projected ring structure is, they point out, explained by Humphreys et al. (2002) as tangential amplification arising from a radial velocity gradient at the inner boundary of the SiO maser region. Their maps also show structures outside this inner boundary including 'spatially coherent arcs and filaments'. Such filaments are an interesting, even characterizing feature of our maps at Epochs III and IV and while their published maps present a well-defined ring with a strong concentration of emission in the CSE, there are fewer filaments of the type clearly seen in our Figs. 6 and 7. However, a close look at the movie presented by DK03 (http://www.journals.uchicago.edu/ApJ/journal/issues/ApJ/v599n2/58473/video2.mpg) does reveal filamentary structures along which the emission appears to propagate outwards to the end of the filament. This is consistent with our data where we see a gradient from high to low velocity along the filaments as they propagate outwards at a slight angle to the normal to our line of sight. They find that expansion is the dominant overall velocity over optical phases of $\phi = 0.7$ to 1.5 and this is observed as well in our maps of both maser lines from Epoch III to IV. We note however that the mean radii of the $v = 2$ maser ring shows a slight contraction (see Sect. 5.2). An interesting difference between our maps and those of the previous cycle (DK03) is the apparent lack of activity in the south-east region. Both $v = 1$ and $v = 2$ masers are relatively weak or even absent where there appeared to be considerable turbulent activity and strong emission earlier.

Turning to the relative properties of the $v = 1$ and $v = 2$ masers, we compare our measurements with the single-epoch VLBA observations of the two maser lines by Desmurs et al. (2000). In their work, although the two transitions were observed simultaneously, the maps of the two masers were constructed independently and aligned using a common velocity, point-like reference feature which was assumed to be spatially coincident in both maps. They see rings of masers and among about 17 maser features identified in both lines they find that, in general, the $v = 2$ maser ring is found inside the $v = 1$ masers. They measure the difference of the ring radii in the two maser maps to be $1.3 \pm 0.2$ mas. Considering that the ring thickness is comparable to the size of maser clumps shown in the maps, we believe that the difference they found would result in a partial displacement throughout the co-existing maser clumps in both lines but not a complete separation between the two rings. This situation is seen clearly in our aligned map at Epoch IV (see Fig. 2). Meanwhile, our aligned map at Epoch I shows both a partial displacement of the two clumps on the south-west side and a complete separation over a partial arc on the north-east side. On the other hand, the aligned map at Epoch III reveals near-coincident features among the co-existing masers in the two lines. Desmurs et al. argued that their result showing the systematic inner location of the $v = 2$ masers suggests a radiative pumping scheme, but given the shifts in relative displacement seen in our multi-epoch study, we would caution that a single epoch result may confuse the interpretation. We suggest that other maser models which are predominantly collisionally pumped may more readily predict the observed ring properties (for details see the following section).

In a more recent paper, Cotton et al. (2004) have measured the $v = 1$ and $v = 2$ maser distributions at several epochs toward some Mira variables, not including TX Cam. Their general results show that the $v = 2$ maser shell is always smaller and less variable than that in the $v = 1$ transition. Nevertheless, an inspection of their diagrams shows that for some objects the $v = 1$ and $v = 2$ lines occur in the same region at some epochs. These properties are similar to our results for TX Cam and mitigate against pure radiative pumping. Contrary to our results Cotton et al. find that for some sources the distribution of $v = 1$ and $v = 2$ masers are quite different.

## 5.2. Comparison with models

The most detailed simulation work published to date on time variable multi-transition SiO emission in CSEs is that from Gray & Humphreys (2000; GH00 hereafter) and Humphreys et al. (2002; H02 hereafter). Both papers are based on the same maser model which is predominantly collisionally pumped, although a component of radiative pumping is included. In the model a pulsation driven shock propagates outward and causes spatial variations in density, temperature, and velocity field and hence in the emissivity in the different maser lines. In contrast the IR field is kept constant throughout the cycle. H02 presents predictions for the cycle-dependant emissivity in the $v = 1$, $J = 1 - 0$ and the $v = 1$, $J = 2 - 1$ transitions. The $v = 1$, $J = 2 - 1$ transition occurs at 3mm wavelength and we have no observations of this line, however we can make detailed comparisons between our $v = 1$, $J = 1 - 0$ observations and the synthetic images they present over a full stellar pulsation cycle. In an earlier paper (GH00), results on the comparison of the $v = 1$ and $v = 2$, $J = 1 - 0$ masers were presented. These results can be compared directly to our dual transition observations.

In comparing our observations with the models for $v = 1$, $J = 1 - 0$ in H02 we must take into account the difference in definitions of the stellar phase of zero. In the model, zero phase was defined as when a pulsation-driven shock front emerges from the stellar photosphere. Observationally the zero of optical phase is defined at maximum optical light. H02 found an empirical relation between the two definitions of phase; the model phase of maser minimum light was compared to the optical phase of the image from the sequence of the VLBA monitoring of TX Cam, which showed the disruption of the maser ring appearing at maser minimum light. From the paper of H02 we deduce using this comparison that optical phase = model phase + 0.22 cycles, and therefore that the corresponding model phase at optical maximum is 0.78. Using the above relationship and the optical phases of our epochs given in Table 2 we can compare our images with the different model epochs in H02. Epoch I should correspond to model epoch $\sim$ 7 in Fig. 5 of H02 and Epoch II corresponds to model epoch $\sim$ 9 which is almost at the model maser minimum (epoch 10). The general dearth and weakness of features predicted near maser minimum is consistent with our data seen at these two epochs. Epoch III corresponds to model epoch $\sim$ 13 and is nearly half way from maser minimum to maximum in the model. Finally, Epoch IV corresponds to model epoch 17, which is at predicted maser maximum, again consistent with the structure observed (see Fig. 6 of H02). The optical phase of Epoch IV of 0.05 shows that as predicted the maser maximum light occurs sometime after the optical maximum light. Observationally, it is well known that there is an average phase lag of the SiO masers (at 43 and 86 GHz) intensity maxima of 0.1-0.2 relative to the visual maxima, but this value is variable between stars and between different cycles of the same star (Alcolea et al. 1999).

Turning to the model predictions for $v = 1$ and $v = 2$, $J = 1 - 0$ presented in GH00 we assume the same relationship between the quoted model phase and observed optical phase. In our observations we find at Epoch III that the ratio of the ring radius (Sect. 4.3) of the $v = 2$ maser to the $v = 1$ maser is 96% while at Epoch IV it is 91%. From Fig. 1 in GH00 the simulations predict the radius ratio to be 94% and 92% at the model epochs corresponding to our Epochs III and IV, showing excellent agreement between our observation and the model. The reason for the smaller relative radius at the later epoch is that, in the model, just after maser maximum there is a slight contraction of the $v = 2$ maser ring radius - while the $v = 1$ maser ring stays almost the same size. These results are exactly mirrored in our observations as described by the radius at 50% emission seen in Sect. 4.3 and shown in the comparison of the insets in Figs. 8 and 9.

As well as predicting the relative radii in the $v = 1$ and $v = 2$, $J = 1 - 0$ masers, with the $v = 2$ maser being inside the $v = 1$ maser the GH00 paper also makes predictions about the range of ring radii or ring thickness in the rings in the two transitions. The model predicts that the $v = 1$ maser ring is always thicker than the $v = 2$ ring. This arises because of the wider range of physical conditions supporting inversion in the $v = 1$ transition. Observationally we find using the 25% and 75% percentiles that the ring thickness in the $v = 1$ and $v = 2$ masers is 15.6% and 14.9% of the radius at Epoch III and 19.5% and 18.6% at Epoch IV. These values suggest only a slightly wider region for the $v = 1$ lines. However looking at Figs. 8 and 9 it appears that the main peak of $v = 1$ is somewhat wider than the main peak of $v = 2$ at both epochs, but when one includes the highest and lowest radius emission the difference is less clear. The intercomparison of the relative ring thickness is clearly sensitive to the exact definition of ring thickness chosen. In fact the GH00 paper estimated that the $v = 1$ maser shell thickness should be double that of $v = 2$ but the statistics used were very different from our statistics based on quartiles. GH00 define the ring thickness as the difference in radius between the outermost and innermost contributing maser spot from the 50 brightest at any phase. However, in this case, looking at the synthetic images of $v = 1$, $J = 1 - 0$ masers in the paper of H02 we notice that the innermost maser can be found from a feature forming on the disk of the star and so this may not be a very useful statistic.

The model of Bujarrabal (1994b) assumes purely radiative pumping and uniform spherical shells at a single phase. This model predicts observed shell widths and radii which can be compared to our observations. Numerical simulations using different parameters gave maser ring widths varying between $2 \times 10^{11}$ cm and $4 \times 10^{13}$ cm. For shell widths smaller than $10^{12}$ cm the maser shells were saturated. Shell radii were between $1 - 2 \times 10^{14}$ cm. Our observations show for Epochs III and IV ring widths of $1 - 2 \times 10^{13}$ cm and ring radii $7 - 8 \times 10^{13}$ cm (see 4.3) and therefore seem consistent with their simulations of the unsaturated case. The model also predicts that spatially compact saturated spots should show a relatively small variability, while extended unsaturated components should vary more. Observationally, from Epoch II to IV our impression is that if anything the compact features are more variable than the extended masers.

As Bujarrabal noted, radiative pumping requires systematically different physical conditions to pump the $v = 1$ and $v = 2$, $J = 1 - 0$ masers. This predicts that for masers pumped purely radiatively there should never be features in the two transitions that are coincident. In fact in Epochs III and IV we see many features that are coincident in space and velocity. We believe that these overlapping features are too numerous and too similar to be an artifact of our alignment process. In particular we find that many features are spatially resolved and show almost exactly the same shape and size in the two transitions. This result is not consistent with the simplest radiative-only pumping models. In contrast the work of Lockett & Elitzur (1992) has shown that collisional pumping is more robust and can produce $v = 1$ and $v = 2$ masers in the same region of space over a broad range of physical conditions.

## 5.3. Radial spokes and the shell kinematics

A striking result in Epochs III and IV in the total intensity images (see Fig. 3) is the presence of radially directly lines or *spoke-like* features. The velocity field at these epochs (see Figs. 4 - 7) shows that these spokes are mainly red-shifted, although a few short blue shifted spoke-like features may also be present. There are clear velocity gradients along these spokes in the sense that the velocity decreases with increasing radius. A related observation is that there are relatively few systemic velocity masers of any morphology. While the exact value of the systemic velocity is probably uncertain to a few km s$^{-1}$ the spectra (see Figs. 4 - 7) also seem to indicate a deficiency of emission in the centre of the spectrum. These observations are surprising because in the standard model of radial outflow and tangential shell amplification the brightest masers should preferentially be at the systemic velocity.

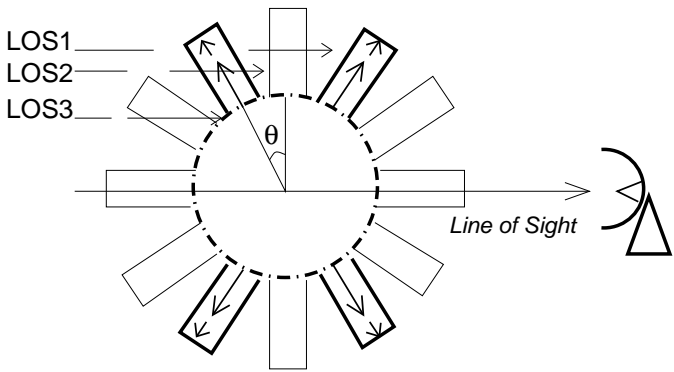

**Fig. 10.** Spoke geometry and possible mechanism for preferentially detecting slightly red- and blue-shifted spokes. The rectangles indicate spokes of gas flowing out from the star at different angles around the shell $\theta$. The thick line rectangles are the brightest spokes which we observe. All spokes have the same velocity field which is decelerating with radius. The line of sight through a spoke in the sky plane (at $\theta = 0$), indicated by LOS1, has a velocity coherent path equal to the spoke width. As we go around the shell in angle $\theta$ the coherent paths through spokes increases proportional to $1/\cos(\theta)$. At some point around the shell, the line of sight through a spoke, indicated by LOS2, has a maximum velocity coherence length. This occurs where the change in velocity along this LOS equals the gas internal velocity dispersion. Further around the shell, such as the spoke crossed by LOS3, the velocity coherent path length decreases because of the large velocity gradient along the path. Slightly red- and blue-shifted spokes, such as those crossed by LOS2, which have the largest coherent paths will be brightest and most easily detected.

The observed spoke-like features might be explained if there are non-radial gas motions caused by turbulence, but in this case it is hard to understand the lack of systemic velocity components or why the size of the spoke velocity gradients are so similar. The observed pattern instead suggests a selection effect which makes slightly red-shifted/blue-shifted spokes more detectable. In this case with a decelerating radial outflow, the observed spoke velocity gradients are then naturally explained (see Fig. 10). Slightly red- and blue-shifted spokes will be more prominent if they have longer coherent velocity path lengths. As we move around the shell the coherent path length through spokes will increase because of the larger geometrical path through the spokes. This growth in path length is however limited by the radial deceleration along the spokes. At some critical angle around the shell $\theta_c$ the velocity difference along the path through a spoke will equal the gas internal velocity dispersion and a maximum velocity coherent path is obtained. Spokes at larger angles will have ever shorter coherent paths set by the internal velocity dispersion and gas radial deceleration rate. The spokes at $\theta_c$ around the shell will be the brightest ones, especially if the maser is saturated. In this case in each spoke there will be strong beaming along the dominant paths which are at an angle $\theta_c$ to the spoke axis.

In its simplest form the above model would predict equally bright blue- and red-shifted spokes, contrary to our observations. This asymmetry can however be explained if there is a radial gradient in the maser source function (see Elitzur 1992) for spontaneously emitted seed photons. Spokes in which the source function was larger at the far end of the path would be brightest, because the seed photons would have a longer path for amplification. Since our red-shifted spokes are more prominent this implies an increasing source function versus radius. Finally this model may also naturally explain the large brightness variations observed along the spokes (see Sect. 4.1). If the spokes are slightly conical rather than cylindrical the angle between the dominant path and the spoke axis will vary along the spoke. This will mean that a spoke will have a brightest point along it where the beaming cone exactly intersects the line of sight. Potentially this mechanism could cause apparent non-physical phase motions between epochs if conical spokes expanded or contracted slightly between epochs.

It is intriguing that our red- and blue-shifted spokes in our maps have similar velocities to the red- and blue-shifted broad wing emission found in single dish observations (Cernicharo et al. 1997; Herpin et al. 1998) of TX Cam and other sources. Herpin et al. noted the importance of obtaining interferometric observations of this broad wing emission, observations which we now have. These authors also noted that the broad wing emission is most prominent near maser maxima which is exactly when we observe the red-shifted/blue-shifted spoke-like features in TX Cam. Cernicharo et al. (1997) suggested that broad wing emission could be ascribed to inflowing or outflowing gas in front of and/or behind the star without explaining how such non-tangential lines of sight can produce bright maser emission. The breaking of spherical symmetry by spokes as described above may provide a suitable mechanism.

It is presently unclear what the observed spokes correspond to physically. They could correspond to real positional variations in gas density due to Kelvin-Helmholtz or other instabilities in the shell. They could represent areas with different pumping conditions. Finally they may simply represent small regions of highly velocity coherent gas in ordered flows which lie in between bubbles of highly turbulent gas.

As well as the spoke velocity gradients a final general conclusion we can draw from the observed velocity field is that blue- and red-shifted emission is present throughout the ring. This rules out rotation in the stellar envelope of TX Cam. There are, however, recent VLBA observations (Boboltz & Marvel 2000; Hollis et al. 2001) suggesting rotation of the SiO maser shells around NML Cyg and R Aqr. Since NML Cyg is a red supergiant and is predicted to become a supernova in the near future and R Aqr is a binary system comprised of a long-period variable which has SiO maser shell and a hot companion/accretion disk, the rotation may well be real. However we caution that in a single epoch our data could be thought to represent rotation, but that the full data on both transitions rules this out in the case of TX Cam (see also DK03).

# 6. Summary and conclusions

We have conducted simultaneous observations of $v = 1$ and $v = 2$, $J = 1 - 0$ SiO masers toward the Mira variable TX Cam using the VLBA at four different stellar phases, cover-ing from maser minimum light to maser maximum light. We used a new technique in order to determine the correct relative positions of the masers in the two lines. Although limited by atmospheric effects and source positions we estimate that this method achieved a relative positional accuracy of 0.2 mas. With a more optimised schedule we believe the technique can in the future achieve an even better accuracy.

Our results allow us for the first time to compare the stellar phase-dependent properties of SiO masers in the $v = 1$ and $v = 2, J = 1-0$ transitions. Our main results can be summarised as follows:

1. In both transitions there is missing diffuse flux density which is not recovered in our maps. Based on the autocorrelation spectra and the flux density observed on the shortest VLBA baselines we estimate that this missing flux is in structures larger than 3 mas.

2. Masers form ring-like structures near the period of maser maximum light. Disruption of the maser rings occurs near to maser minimum light. The largest extent of the masers in both transitions shows expansion over three consecutive Epochs from II to IV. In contrast the median radius of the rings first expands but then between Epochs III and IV is stationary in the $v = 1$ maser ring or slightly contracts in the $v = 2$ maser ring.

3. The $v = 1$ maser ring appears to be somewhat thicker than the $v = 2$ maser ring, however the extent of the difference depends on the statistics used to quantify ring thickness. Statistics which concentrate on the brighter regions give a slightly larger difference in thickness, while if the ring thickness is defined by its more extended weak emission the difference is less.

4. Aligned maps of the two masers show the $v = 2$ masers residing slightly inside $v = 1$ masers in general but the mean radial displacement of the two masers changes with stellar phase. In addition to this general statistical trend, at Epochs III and IV we find that several individual maser features overlap almost exactly in both position and velocity.

5. With respect to the maser pumping mechanisms, the result of point 4 seems to rule out purely radiative pumping models. Other stellar phase-dependent properties of the two masers found in our four epoch observations are more consistent with the maser model whose physical conditions are mostly determined by stellar pulsation-driven shocks.

6. The observed velocity field in the two transitions shows that both the red- and blue-shifted emission arise in all parts of the ring but red-shifted components are brighter than blue-shifted ones at all four epochs. At all epochs the red-shifted components are more dominant in the $v = 2$ masers compared to the $v = 1$ masers. There seem to be relatively few bright maser features at the systemic velocity.

7. We observe spoke-like features around the ring in both transitions, particularly in Epoch IV, which have clear radial velocity gradients. These gradients imply a decelerating outward flow velocity with increasing radius. These observations and the relative lack of systemic masers argues against a model with pure tangential amplification in spherical shells. We speculate that the presence of spoke-like structures may be enough to break the spherical symmetry and preferentially select slightly red-shifted and blue-shifted masers.

*Acknowledgements.* We would like to thank the VLBA scheduling, correlator, and operations staff. We acknowledge with thanks data from the AAVSO International Database based on observations submitted to the AAVSO by variable star observers worldwide.

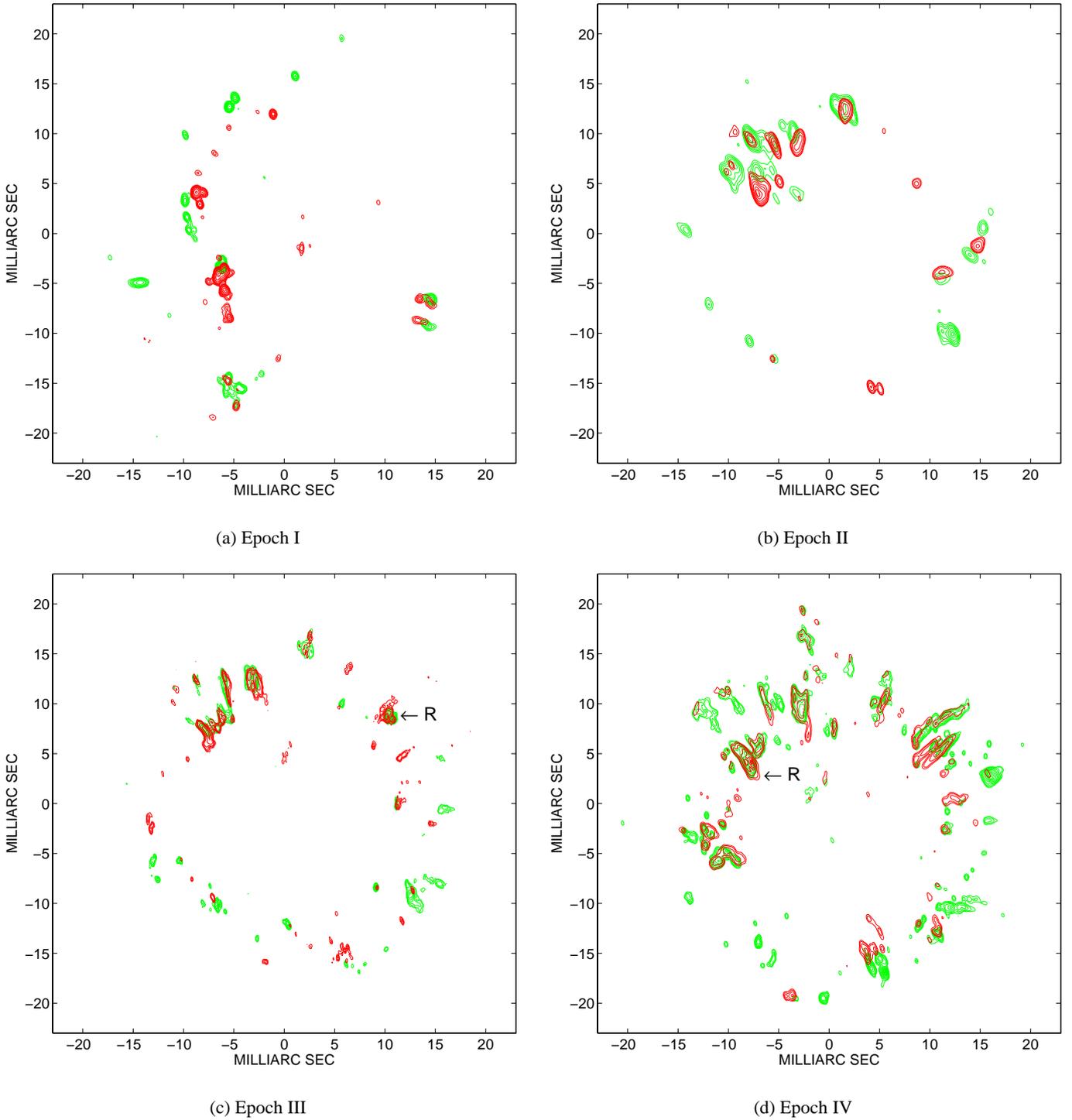

(a) Epoch I

(b) Epoch II

(c) Epoch III

(d) Epoch IV

**Fig. 2.** The $v = 1$ contour (green) maps aligned with $v = 2$ contour (red) maps at four epochs. In Epochs III and IV the 'R' indicates the feature used for registration. Epoch I : For $v = 1$ contours cover irregularly the range between 0.38 - 10.98 Jy/beam and for $v = 2$ from 0.79 - 29.13 Jy/beam. Epoch II : For $v = 1$ contours cover the range between 0.83 - 7.85 Jy/beam and for $v = 2$ from 1.66 -18.65 Jy/beam. Epoch III : For $v = 1$ contours cover the range between 0.85 - 39.98 Jy/beam and for $v = 2$ from 0.90 - 75.06 Jy/beam. Epoch IV : For $v = 1$ contours cover the range between 1.62 - 80.75 Jy/beam and for $v = 2$ from 1.90 - 94.93 Jy/beam.

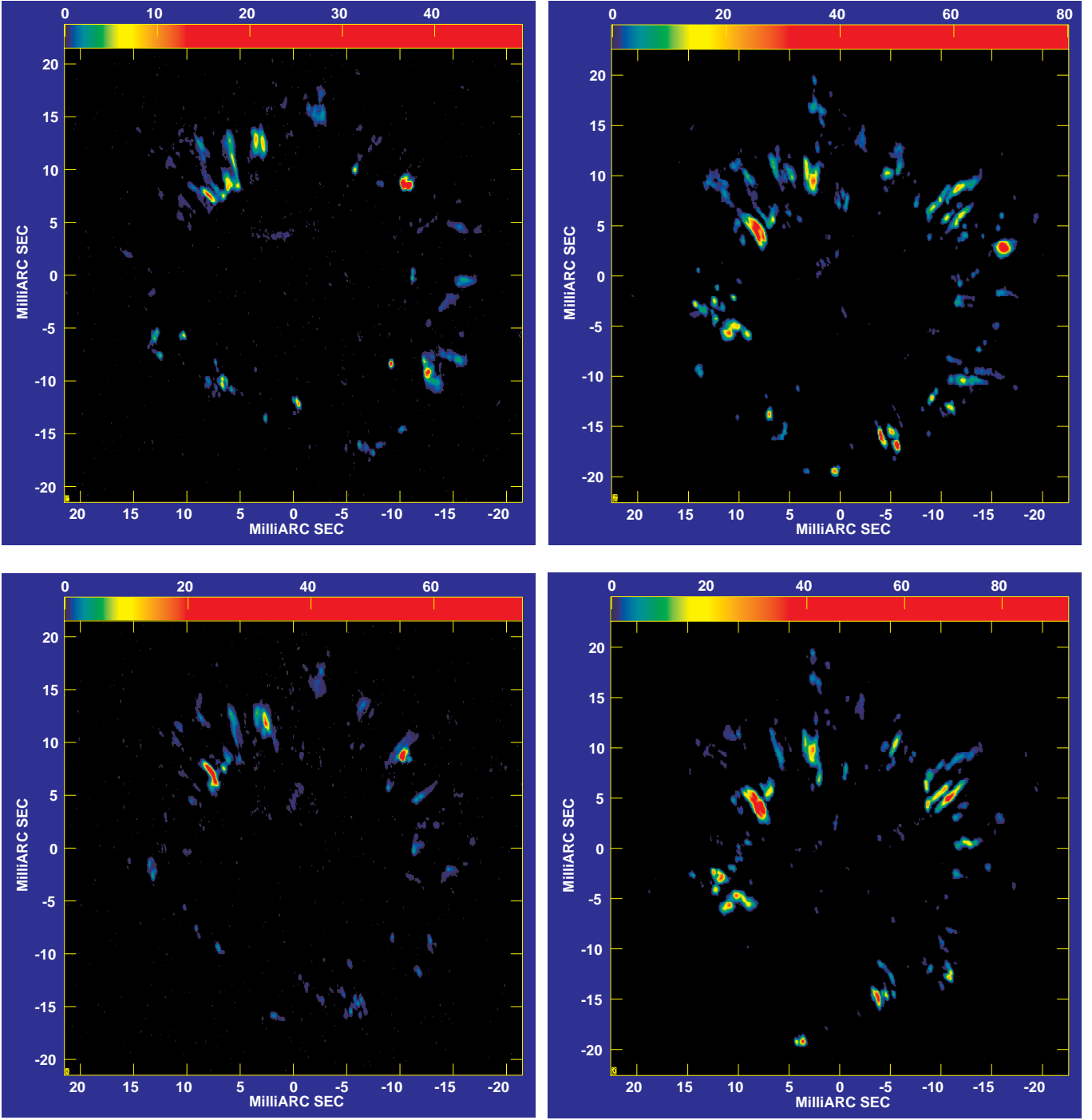

**Fig. 3.** VLBI maps of $v = 1$, $J = 1 - 0$ (upper) and $v = 2$, $J = 1 - 0$ (lower) SiO maser emission toward TX Cam, summed over all channels of emission, observed at Epoch III (left) and at Epoch IV (right). The colour bar shows the flux scale in Jy/beam

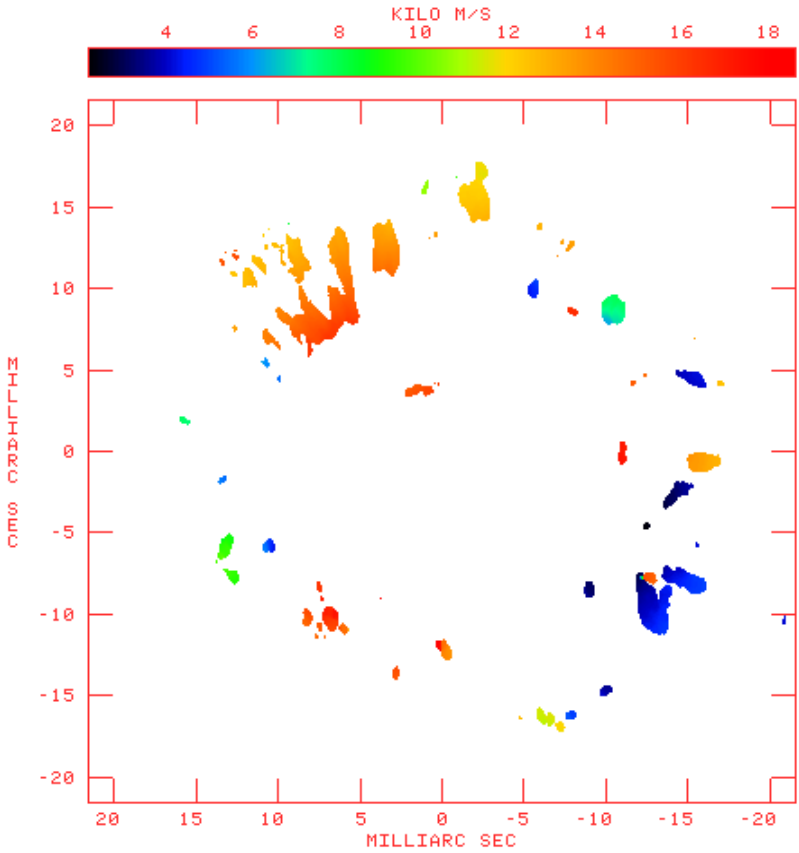

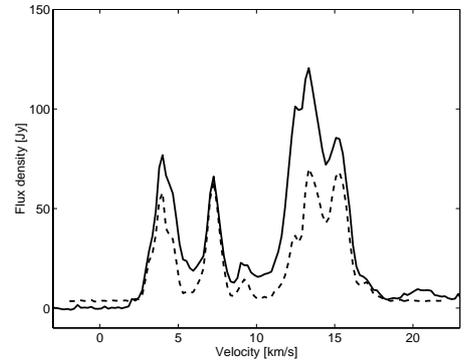

**Fig. 4.** Left : Velocity field of the masers in the $v = 1$ line of TX Cam observed at Epoch III. Horizontal colour bar gives mean LSR velocity at each position. Compare to Fig. 3 top left for the corresponding integrated intensity map. Right: Spectrum of the integrated flux density from all the maser emission in the map (dashed line) together with the total power spectrum (solid line).

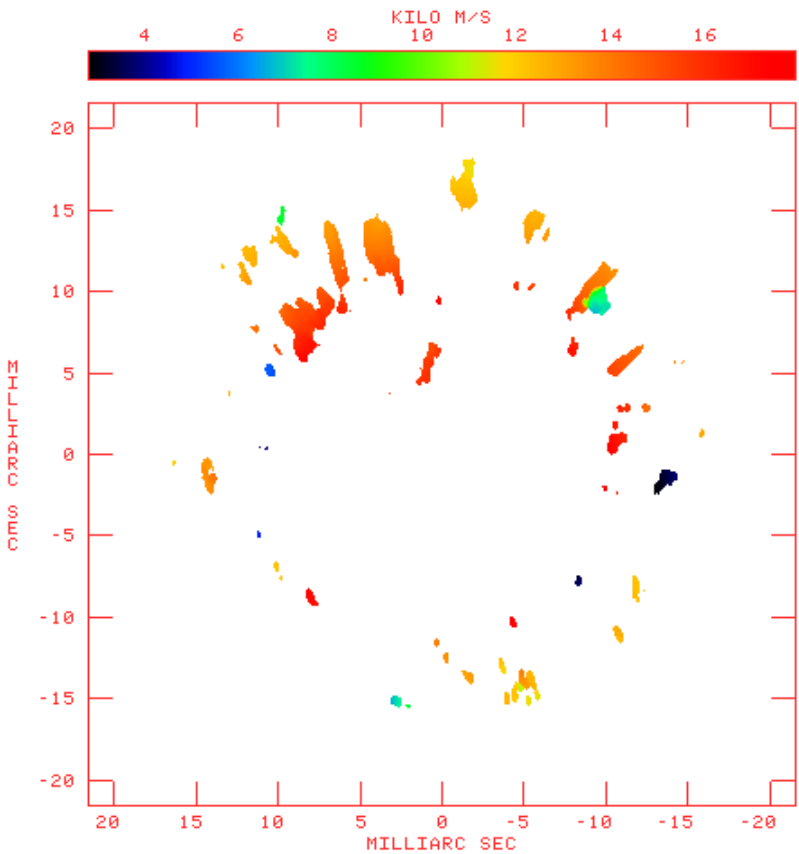

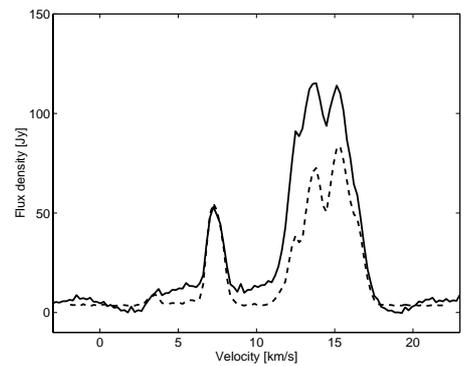

**Fig. 5.** Left : Velocity field of the masers in the $v = 2$ line of TX Cam observed at Epoch III. Horizontal colour bar gives mean LSR velocity at each position. Compare to Fig. 3 bottom left for the corresponding integrated intensity map. Right: Spectrum of the integrated flux density from all the maser emission in the map (dashed line) together with the total power spectrum (solid line).

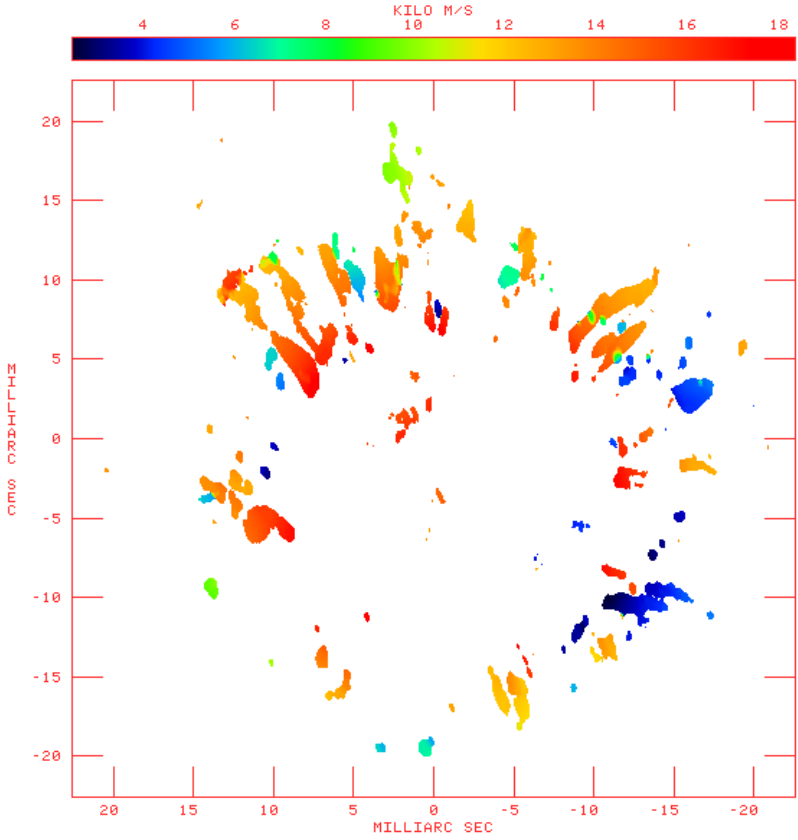

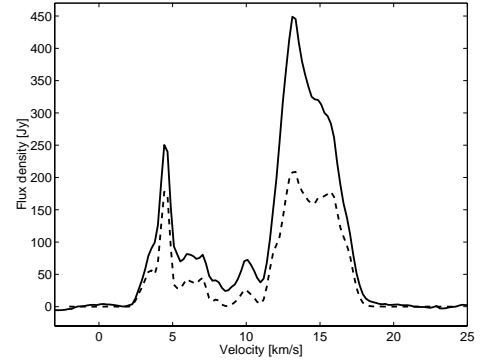

**Fig. 6.** Left: Velocity field of the masers in the $v = 1$ line of TX Cam observed at Epoch IV. Horizontal colour bar gives mean LSR velocity at each position. Compare to Fig. 3 top right for the corresponding integrated intensity map. Right: Spectrum of the integrated flux density from all the maser emission in the map (dashed line) together with the total power spectrum (solid line).

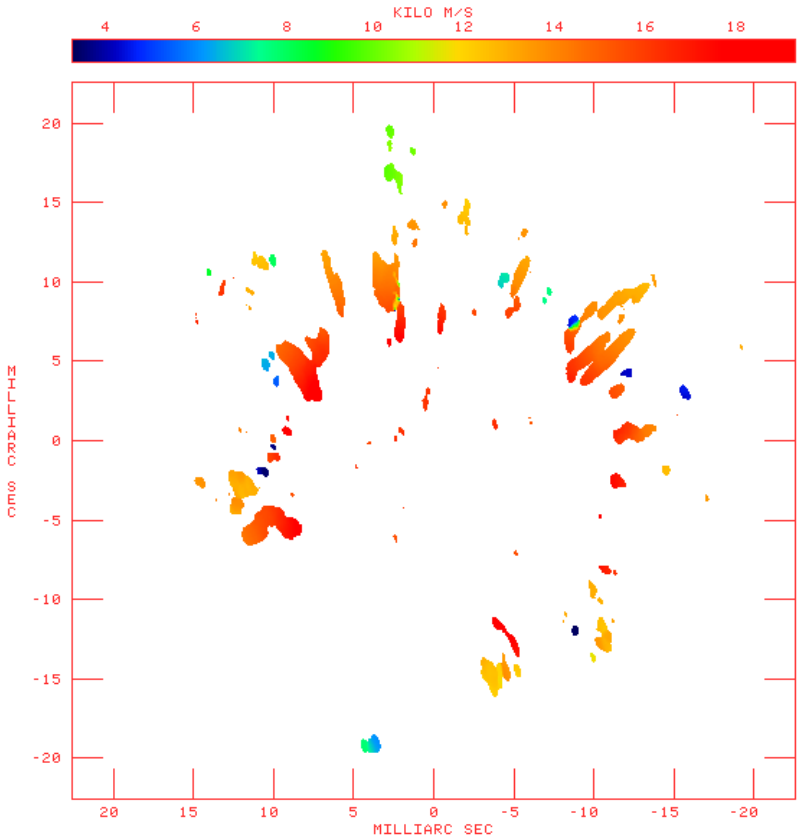

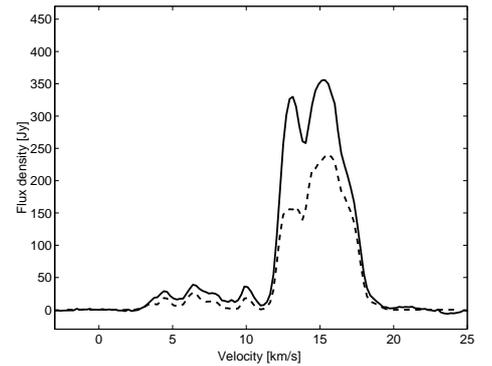

**Fig. 7.** Left: Velocity field of the masers in the $v = 2$ line of TX Cam observed at Epoch IV. Horizontal colour bar gives mean LSR velocity at each position. Compare to Fig. 3 bottom right for the corresponding integrated intensity map. Right: Spectrum of the integrated flux density from all the maser emission in the map (dashed line) together with the total power spectrum (solid line).